\documentclass[prd,showkeys,floatfix,nofootinbib,
               preprint,12pt,
               tightenlines,fleqn]{revtex4-1}

\usepackage{amsmath,amssymb,revsymb,graphicx,dcolumn}
\usepackage{leftidx}
\usepackage{slashed}

\newcommand{\beq}{\begin{equation}}
\newcommand{\eeq}{\end{equation}}
\newcommand{\beqa}{\begin{eqnarray}}
\newcommand{\eeqa}{\end{eqnarray}}
\newcommand{\bsubeqs}{\begin{subequations}}
\newcommand{\esubeqs}{\end{subequations}}

\newcommand{\PP}{p}
\newcommand{\KK}{k}
\newcommand{\PPprime}{p'}
\newcommand{\KKprime}{k'}

\usepackage{mwe}
\newcommand{\imineq}[2]{\vcenter{\hbox{\includegraphics[height=#2ex]{#1}}}}

\begin{document}
\noindent Class. Quant. Grav. \textbf{35}  (2018) 125004
\hfill   arXiv:1710.06405\;  %%(\version)
%
%arXiv:1710.06405 \hfill  KA--TP--33--2017\;(\version)
%
%arXiv:xxxxxxxxxx
%\preprint{KA--TP--33--2017\,(\version)}
%
\newline\vspace*{3mm}

\title[]{Across-horizon scattering and information transfer}

\author{V.A. Emelyanov}
\email{viacheslav.emelyanov@kit.edu}
\affiliation{Institute for
Theoretical Physics, Karlsruhe Institute of
Technology (KIT), 76128 Karlsruhe, Germany\\}  %%FRK

\author{F.R. Klinkhamer}
\email{frans.klinkhamer@kit.edu}
\affiliation{Institute for
Theoretical Physics, Karlsruhe Institute of
Technology (KIT), 76128 Karlsruhe, Germany\\}

\begin{abstract}
\vspace*{2.5mm}\noindent
We address the question whether or not two electrically charged
elementary particles can Coulomb scatter if one of these particles is
inside the Schwarzschild black-hole horizon and the other outside.
It can be shown that the quantum process is
consistent with the local energy-momentum conservation law.
This result implies that across-horizon scattering is a physical effect,
relevant to astrophysical black holes.
We propose a \textit{Gedankenexperiment}
which uses the quantum scattering process to
transfer information from inside the black-hole horizon to outside.
\end{abstract}


\maketitle

\section{Introduction}
\label{sec:Introduction}

Astrophysical black holes are expected to evaporate~\cite{Hawking1975}.
This quantum process leads to the information-loss problem in
black-hole physics~\cite{UnruhWald2017}.
A final resolution of this problem could, as sometimes suggested,
be based on principles that lie outside the framework
of the standard semi-classical physics.
But perhaps these nonstandard principles are not really needed,
as the high-energy photon of quantum electrodynamics
(QED) may be capable of propagating out of the black-hole horizon~\cite{Emelyanov2017}.

In this paper, we implement a particular approach to
partial information recovery out of a black hole,
without changing the principles of local quantum field theory.
The main idea is based on the circumstance that
a quantum field simultaneously exists at all
spacetime points, including those of the region
inside the black-hole horizon. Specifically, a
quantum field is nonvanishing at each point of a
given spacetime manifold and, according to the
canonical (anti-)commutation relations, gives rise
to random fluctuations of observables at each
spacetime point, even in the vacuum state.
A particle is a localized excitation, which significantly disturbs
the quantum-field expectation value in its neighborhood.
If the particle is sufficiently close to the black-hole horizon,
the corresponding quantum-field disturbance is nonvanishing even across the
black-hole horizon. Since the quantum-field disturbance is not a quantity
that must propagate at a particular speed (unlike real particles),
this may cause the following across-horizon effect:
two charged elementary particles, located at different sides of the black-hole horizon,
can scatter with each other via their quantum-field disturbances.

More concretely, the across-horizon-scattering effect
can be understood as follows.
In QED, the scattering interaction between charged elementary
particles at tree level
(virtual-photon exchange) is mathematically described
by the Feynman propagator $G_{F}^{\,\mu\nu}(x,x')$ of the electromagnetic field
(and, at higher orders in perturbation theory,
also by the Feynman propagators of the electrically charged matter
fields).
This particular Green's function
has a nonvanishing support for spacelike-separated points:
$G_{F}^{\,\mu\nu}(x,x') \neq 0$ holds even
for two spacetime points with spacelike separation,
whereas, for example, the retarded Green's function
$G_{R}^{\,\mu\nu}(x,x')$ vanishes identically for
two spacetime points with spacelike separation.
Hence, QED does not forbid the interaction between two
electrically charged particles that are causally disconnected.

The crucial question, now, is if  across-horizon
scattering can be used to transfer information from
the inside of the black-hole horizon to the outside.
The present paper shows that, in practice,
it may be difficult to use the across-horizon-scattering effect
for information transfer but it is not impossible.

We recently became aware of an earlier calculation~\cite{Reznik-etal-2005}
of the emerging entanglement of two spatially separated detectors
with interactions from a relativistic scalar field
(see, e.g., Ref.~\cite{Pozas-Kerstjens-Martin-Martinez2016}
for further references).
The setup of Ref.~\cite{Reznik-etal-2005} is similar to ours
and we will comment on it in
Sec.~\ref{sec:Discussion}.

Throughout this article, we take
$c = G_{N} = \hbar = 1$, unless otherwise stated.
Gravity is assumed to be described by
standard general relativity, based on the Einstein Equivalence Principle,
and the metric signature is $(+\,-\,-\,-)$.
Interactions of elementary particles are taken to be described by
standard quantum field theory over curved spacetime.

%%\newpage%%tmp
\vspace*{0mm}
\section{Scattering}
\label{sec:Scattering}
\vspace*{0mm}

It appears that the Universe can be locally
approximated by Minkowski spacetime.
The Minkowski spacetime manifold is a fundamental ingredient of
elementary particle physics  for the description of high-energy
interaction processes. In QED, the leading-order
probability amplitude for the Coulomb scattering of two
electrically charged elementary particles
(electron $e^{-}$ and muon $\mu^{-}$)
is represented by the following tree-level
momentum-space  %%FRK
Feynman diagram~\cite{Feynman1949,%
ItzyksonZuber1980,Veltman1994,PeskinSchroeder1995,Schwartz2014}:
\beqa
\label{eq:2-2-scattering}
\hspace{-4mm}{}_\text{out}
\langle\mu,\,\PPprime\,;\,e,\,\KKprime| \mu,\,\PP\,;\,e,\,\KK\rangle{}_\text{in}
&\propto&
\mathbf{\imineq{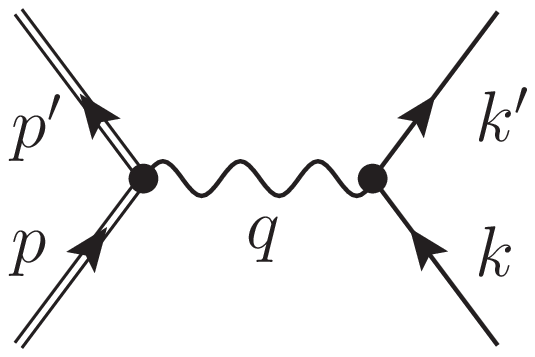}{11.0}}
\hspace{-4mm}\,,
\eeqa
with definition $q \equiv \PPprime - \PP$
and arrows showing the flow of negative electric
charge (and also the flow of positive lepton number,
$L_{\mu}=1$ for the $p$ and $p'$ particles      %%FRK
and $L_{e}=1$ for the $k$ and $k'$ particles).  %%FRK
The wavy line in the diagram on the right-hand-side of
\eqref{eq:2-2-scattering} stands for the
Feynman photon propagator: $-i g_{\mu\nu}/(q^2 +i\epsilon)$
with a positive infinitesimal $\epsilon$.
The Feynman propagator, different from the retarded propagator,
is a crucial ingredient for obtaining a unitary $S$-matrix
(see, e.g., Sec.~3.7 and Chap.~8 of Ref.~\cite{Veltman1994}).  %%FRK
The amplitude \eqref{eq:2-2-scattering} is identically zero, unless
the energy-momentum conservation law is fulfilled,
$\PP+\KK=\PPprime+\KKprime$.

\begin{figure}[t]
\includegraphics[width=8cm,height=4cm]
{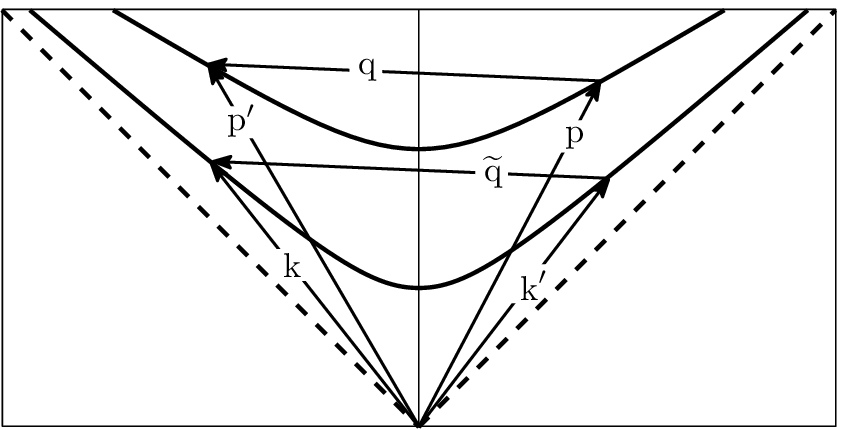}
%Fig1-v7.eps=Fig1-v5.eps
%{space-time-like-vectors-v4035.eps}
%{space-time-like-vectors-v401.eps}
%{space-time-like-vectors-v307.eps}
%{space-time-like-vectors-v29913.eps}
%{space-time-like-vectors-v181.eps}
%{space-time-like-vectors-v073.eps}
\vspace*{0mm}
\caption{Initial-particle momenta ($p$ and $k$)
and final-particle momenta ($p'$ and $k'$),
where the $p$ and $p'$ particles have a larger mass
than the $k$ and $k'$ particles.
Shown is a two-dimensional slice of the four-dimensional
energy-momentum space, with energy along the vertical axis.
The dashed line corresponds to the light-cone and all
four particle momenta are timelike.
Coulomb scattering \eqref{eq:2-2-scattering}
occurs if $q \equiv \PPprime - \PP$
equals $\widetilde{q} \equiv \KK - \KKprime$,
which corresponds to energy-momentum conservation.
}
\label{fig:1}
\end{figure}

In standard QED over Minkowski spacetime, the momenta of the particles
before and after the Coulomb scattering are timelike four-vectors,
whereas the momentum-exchange vectors $q\equiv \PPprime - \PP$
and $\widetilde{q}\equiv \KK - \KKprime$
are spacelike four-vectors.
Figure~\ref{fig:1} gives a concrete example of these
momenta, provided $p'$ and $k'$ satisfy the energy-momentum-conservation
condition ($q=\widetilde{q}$).

%%EqFig12345-v7.eps --> EqFig12345-v8.eps
%%Fig678-v7.eps     --> Fig678-v8.eps

\begin{figure}[t]
\includegraphics[width=8.0cm,height=6.4cm]
{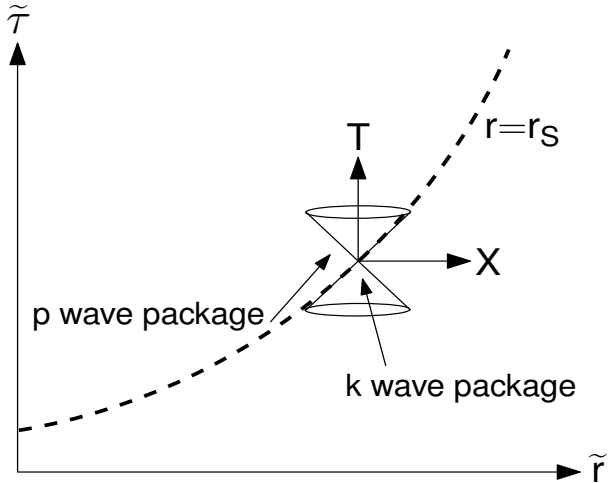}
%{event-horizon-lightcone-v6021.eps}
%{event-horizon-lightcone-v4711.eps}
%{event-horizon-lightcone-u-v305.eps}
%{event-horizon-lightcone-u-v29915.eps}
%{event-horizon-lightcone-u-v103.eps}
%{event-horizon-lightcone-u-v061.eps}
\vspace*{-0mm}
\caption{In the vicinity of the black-hole horizon,
coordinates can be chosen which are locally Minkowskian.
A local inertial coordinate system (coordinates $T,\,X,\,Y,\,Z$)
can, for instance,
be embedded in the $(\widetilde{\tau},\,\widetilde{r}\,)$-plane, where
$\widetilde{\tau}$ and $\widetilde{r}$ are, respectively,
Novikov's infalling-clock time coordinate
and infalling-clock comoving radial coordinate~\cite{Gautreau1995}.
With these local inertial coordinates ($Y$ and $Z$ pointing out
of the plane shown),  the figure sketches the light-cone
in the neighborhood of the Schwarzschild horizon (dashed line)
at standard radial coordinate
$r=r_{S}\equiv 2G_{N} M/c^2$. In addition,
the trajectories of two colliding wave packages are shown,
one wave package with average momentum $k$ is
positioned outside the black-hole horizon
and the other wave package with average momentum $p$
is positioned inside the horizon. These momenta $k$ and $p$
correspond to those of the scattering process
\eqref{eq:2-2-scattering} and also appear in Fig.~\ref{fig:1}.
For ultrarelativistic particles, the momenta $k$ and $p$
are close to their respective light-cones.
}
\label{fig:2}
\end{figure}

Next, consider a neutral and nonrotating black hole of
astrophysical size (mass $M \gtrsim M_\text{Sun}$),
so that the curvature length scale near the horizon is large
compared to the Coulomb-scattering length scale
to be determined shortly.
In the region near the
black-hole horizon, one can always introduce local Minkowski coordinates,
according to the Einstein Equivalence Principle.
For these local inertial coordinates
($T,\,X,\,Y,\,Z$), part of the horizon of the black hole
coincides locally with part of the light-cone shown in Fig.~\ref{fig:2}
(which is a simplified version of Fig.~3(a) in Ref.~\cite{Gautreau1995}).
Locally, there is nothing in the inertial frame
which marks the position of the projected  horizon surface;
see the second paragraph of App.~\ref{app:Initial-particle-trajectories}
for further discussion.
Remark that the black-hole metric for Novikov coordinates
(as used in Fig.~\ref{fig:2}) is nonsingular at $r=R_{S}$;
see, e.g., Sec.~31.4 in Ref.~\cite{MisnerThorneWheeler1973}.
Figure~\ref{fig:2}
also shows the trajectories of two colliding wave packages,
one outside the black-hole horizon with average momentum $k$ and
the other inside the horizon with average momentum $p$.

We are, now, ready to consider the collision of
two electrically charged elementary particles of different mass,
where the muon with mass $m_{\mu}$ is inside the black-hole horizon
and the electron with mass $m_{e}$ is outside.
In a local inertial coordinate system (LICS)
as depicted in Fig.~\ref{fig:2},
both inside-particle and outside-particle
have timelike four-momenta, denoted $\PP$ and $\KK$, respectively.
These particles can interact with each other
by ``exchange of a virtual photon''
if and only if the local energy-momentum conservation law is fulfilled, namely, the condition
$\PP+\KK=\PPprime+\KKprime$ must hold (Fig.~\ref{fig:1}).
In addition, the initial particles are arranged to
have their closest approach
(proper distance $d=d_\text{min}$ at $T=T_\text{min}$)
near the event horizon, with the muon inside and the electron outside.
Note that, strictly speaking,
the minimum proper distance $d_\text{min}$
is to be determined from the initial-particle trajectories
at $T\ll T_\text{min}$, when interaction effects are negligible.
An example of this setup is given
in App.~\ref{app:Initial-particle-trajectories}.

The across-horizon Coulomb scattering (AHCS) with these
initial momenta and trajectories has
a nonvanishing probability in quantum electrodynamics,
\beqa\label{eq:P-AHCS}
\hspace{-2mm}
P_\text{AHCS}\,\Big|^\text{LICS}
&=&
\left(\;\left|
\mathbf{\imineq{across-horizon-scattering-Eq1-v8.eps}{11.0}}
%\mathbf{\imineq{coulomb-scattering-v307.eps}{11.0}}
%\mathbf{\imineq{coulomb-scattering-v0533.eps}{11.0}}
\hspace{-4mm}\right|^2\;\right)_
{\begin{minipage}{5.5cm}
$r(\PP\,\text{wave package})\big|_{T \leq T_\text{min}}<r_S$
\\[1mm]
$r(\KK\,\text{wave package})\big|_{T \leq T_\text{min}}>r_S$
\end{minipage}}^{\PP+\KK=\PPprime+\KKprime}
> \; 0\,,
\eeqa
where the standard radial coordinate $r$ on the right-hand side
is expressed in terms of the local inertial coordinates
$\{T,\,X,\,Y,\,Z\}$ (cf. Fig.~\ref{fig:2})
and $T_\text{min}$ is the time
when the separation of the initial wave packages
has a minimal proper distance $d_\text{min}$.
The probability \eqref{eq:P-AHCS} will be significant if the minimal
separation $d_\text{min}$ of the two initial
particles is of the order of the
root of the flat-spacetime cross section
(with an infrared cutoff on $q^2$ determined by the experimental setup).

The initial particles of the scattering reaction \eqref{eq:2-2-scattering}
are distinguishable:
the $p$ particle has mass $m_{\mu}$ and lepton numbers
$(L_{e},\,L_{\mu})=(0,\,1)$
and
the $k$ particle has mass $m_{e}$ and lepton numbers
$(L_{e},\,L_{\mu})=(1,\,0)$.
Hence, only the $t$-channel diagram contributes to \eqref{eq:P-AHCS}
and we get the following expression for the flat-spacetime differential
cross section in the ultrarelativistic limit
($E^2 \gg m_{\mu}^2 \gg m_e^2$):
\beq\label{eq:dsigma/dOmega}
\left. \frac{d\sigma}{d\Omega}\,\right|_\text{CM,\,ultrarel}^\text{LICS}
=
\frac{\alpha^2}
     {2\,\left(E_\text{CM}\right)^2\,\left(1-\cos\theta_\text{CM}\right)^2}\,
\Big[4+\left(1+\cos\theta_\text{CM}\right)^2\Big]\,,
\eeq
with $E_\text{CM}$ and $\theta_\text{CM}$, respectively,
the energy of the initial particles
and the scattering angle in the center-of-mass (CM) frame.
In fact, expression \eqref{eq:dsigma/dOmega}
can also be found as Eq.~(5.65) in Ref.~\cite{PeskinSchroeder1995}.
The minimal separation $d_\text{min}$
of the two initial particles in \eqref{eq:P-AHCS}
must then be as close as possible to
\beq\label{eq:d-optimal}
d_\text{optimal}
\sim \alpha\,\hbar c/E_\text{CM}\,,
\eeq
where we have used the inequality
$d\sigma/d\Omega \geq \alpha^2\big/\big(2E_\text{CM}^2\big)$
from \eqref{eq:dsigma/dOmega}
and have temporarily restored $\hbar$ and $c$
($\alpha\approx 1/137$ is the fine-structure constant).
In Fig.~\ref{fig:4} of App.~\ref{app:Initial-particle-trajectories},
we have given an example of initial-particle trajectories
with a finite value of the minimal separation ($d_\text{min}$)
between the initial particles.
But, in principle, it is also possible to consider head-on
collisions.

\begin{figure}[t]
\includegraphics[scale=1.20]
{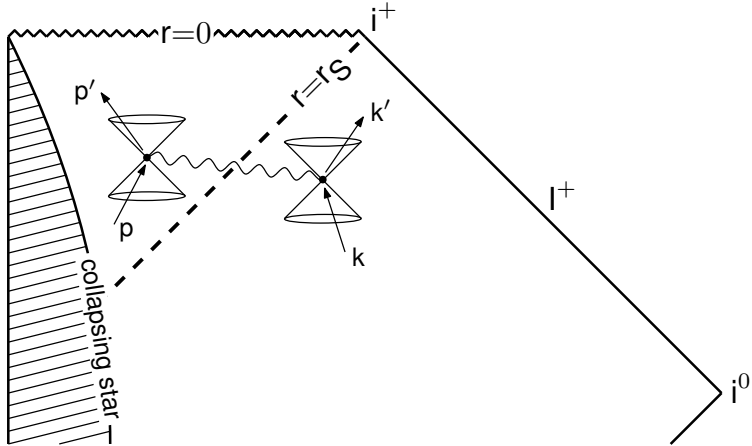}
%Fig3-v7.eps=Fig3-v6.eps
%{virtual-photon-exchange-v50131.eps}
%{virtual-photon-exchange-v4033.eps}
%Fig3-v4.eps \equiv Fig3-v3.eps
%{virtual-photon-exchange-v29921.eps}
\vspace*{-0mm}
\caption{Part of the Penrose conformal diagram
for the Schwarzschild black hole in Kruskal--Szekeres
coordinates (the full diagram is shown as Fig.~24\,(ii)
on p.~154 of Ref.~\cite{HawkingEllis1973},
which contains further details).
Also shown are two light-cones near the
Schwarzschild horizon $r=r_S$ and
the trajectories of two colliding wave packages,
one wave package with average momentum $k$  for the electron is
positioned outside the horizon
and the other wave package with average momentum $p$  for the muon
is positioned inside the horizon.
These elementary particles with momenta $k$ and $p$ scatter
by virtual-photon exchange
(symbolically indicated by the wavy line) and produce, with small
but nonvanishing probability, particles with
momenta $k'$ and $p'$.
The momenta shown correspond to those of the scattering process
\eqref{eq:2-2-scattering}.
For ultrarelativistic particles, the momenta
are close to their respective light-cones
and there is a nonvanishing probability for significant Coulomb
scattering if the $k$ and $p$ wave packages approach each other
within a proper distance of order $\alpha\,\hbar c/E_\text{CM}$,
where $\alpha$ is the fine-structure constant and
$E_\text{CM}$ the center-of-mass energy in the local
inertial coordinate system.
}
\vspace*{0cm}
\label{fig:3}
\end{figure}

%%\newpage%%tmp
The result \eqref{eq:d-optimal} for the optimal
proper distance of the two colliding wave packages
guarantees having a significant scattering probability,
but most scattering will
be in the forward direction, $\theta_\text{CM} \sim 0$.
Thus, in the most probable case, the scattered
outside-particle disappears behind the Schwarzschild
horizon. Still, there is a nonvanishing probability
that the outside-particle (electron) recoils due to the Coulomb
interaction with the inside-particle (muon).
This situation is sketched in Fig.~\ref{fig:3} and
details are given in Apps.~\ref{app:Final-particle-trajectories}
and \ref{app:Probability-amplitude-in-position-space},
with an example of final-particle trajectories
given in Fig.~\ref{fig:5}.
The ultrarelativistic recoil electron $k'$ in Fig.~\ref{fig:3}
initially   %%FRK
crosses the constant-$r$ curves in the Penrose diagram
(cf. Fig.~24\,(ii) in Ref.~\cite{HawkingEllis1973})
and moves away from the black-hole horizon.
There is, then, a nonvanishing probability
that the $k'$ electron triggers an outside-region
detector relatively near to the interaction point.
This last observation is an essential ingredient
of the \textit{Gedankenexperiment} to be discussed
in Sec.~\ref{sec:Gedankenexperiment}.

At this moment, it may be useful to summarize
the three main steps by which we arrived at this surprising result
for the recoil electron:
\begin{enumerate}
\item
According to the Einstein Equivalence Principle,   %%FRK
the physics
in a local inertial coordinate system (LICS) is the \emph{same} as that
of flat Minkowski spacetime, an example being leading-order muon-electron
scattering with the Feynman propagator of the photon
(Feynman propagators are needed for the correct description
of the quantum theory, giving, for example, the correct value of the
Lamb shift; see the discussion in, e.g., Sec.~24.1.2 of
Ref.~\cite{Schwartz2014}).
\item
The previous observation  holds for a freely-falling
LICS at \emph{any} point
of the Schwarz\-schild spacetime manifold outside the singularity,
also near the horizon (part of which coincides with part of the light-cone 
in the LICS; cf. Fig.~\ref{fig:2}).
\item
In the near-horizon LICS, muon-electron \emph{scattering} happens for an
initial muon inside the black-hole horizon
and an initial electron outside the horizon,
with a \emph{nonvanishing} probability for obtaining a recoil
electron in the exterior region
(cf. Fig.~\ref{fig:3} and App.~\ref{app:Probability-amplitude-in-position-space}).
\end{enumerate}
We now turn to an ``application'' of across-horizon Coulomb scattering.

%%\newpage%%tmp
\vspace*{0mm}
\section{Gedankenexperiment}
\label{sec:Gedankenexperiment}
\vspace*{0mm}

The setup of our \textit{Gedankenexperiment} is as follows:
\begin{enumerate}
\item
A large-mass nonrotating black hole
is formed by the spherical collapse of matter and
the resulting spacetime is approximately given
by the static Schwarzschild metric.
\item
After the black-hole formation,
two experimenters,
Castor and Pollux, take up their initial positions outside
the Schwarzschild horizon.
\item
Castor and Pollux intend to use
the across-horizon scattering process from Sec.~\ref{sec:Scattering}
and Fig.~\ref{fig:3} with
pre-determined initial momenta $\PP$ and $\KK$
(in the freely-falling frame), which are
arranged to give, with largest possible probability,
a nontrivial scattering event with recoil momentum $\KKprime$
of the outside-electron.
They also intend to perform their experiments rapidly enough, so that
their later inside/outside positions with respect to the Schwarzschild horizon do not change substantially.
\item
Castor and Pollux, while in the exterior region,
agree on the details of the procedure and get started:
\begin{enumerate}
  \item[4a.]
along a fixed radial direction, Pollux stays outside
the Schwarzschild horizon and Castor rapidly moves inside the horizon;
once inside, Castor must act fast as his available
proper time $\tau^{C}$ is limited,
$\tau^{C} \leq (\pi/2)\,r_S/c \sim 10^{-5}\,\text{s}\,\left(M/M_\text{Sun}\right)$,
see, e.g., Ex.~31.4 on p.~836 in Ref.~\cite{MisnerThorneWheeler1973};
  \item[4b.]
at a pre-arranged moment, Pollux starts to emit,
at regular time intervals, appropriate electrons
[each of momentum $\KK$ from \eqref{eq:2-2-scattering}
and App.~\ref{app:Initial-particle-trajectories},
in the freely-falling frame]
and sends out a finite number $N$ of electrons
in total ($1\ll N < \infty$);
\item[4c.]
starting from the corresponding pre-arranged moment
and at an  appropriately adjusted rate,
Castor either emits $N$ appropriate muons
[each of momentum $\PP$ from \eqref{eq:2-2-scattering}
and App.~\ref{app:Initial-particle-trajectories},
in the freely-falling frame]
or sends no such muons at all;
in the first case, Castor writes in his message-book a ``yes''
and, in the second case, he writes a ``no'';
\item[4d.]
if Castor has emitted $N$ muons $\PP$, then Pollux's detector
(positioned at an appropriate distance away from
the black-hole horizon) has a nonzero chance to register
a momentum change of the exterior electron
as discussed in Apps.~\ref{app:Final-particle-trajectories}
and \ref{app:Probability-amplitude-in-position-space}
[Pollux writes ``1'' in his log-book
if he registers a recoil electron with momentum
$\KKprime$ and ``0''
if he does not register a recoil electron],
but if Castor has emitted no muons $\PP$ at all,
then Pollux's detector will never register a recoil electron
[Pollux writes $N$ times a ``0'' in his logbook];
  \item[4e.]
after $N$ measurements, Pollux looks at his logbook
and summarizes his results as follows:
a sequence of $N$ zeros is written as ``NO''
and a sequence with at least a single ``1''
is written as ``YES.''
\end{enumerate}
\item
According to points 4c and 4e,
Castor can send a message (``yes'' or ``no'' in his message-book),
which is read by Pollux  (``YES'' or ``NO'' in his logbook).
\item
As Castor's message-book is in the interior region of the
Schwarzschild black-hole horizon
and Pollux's log-book in the exterior region,
information (``yes'' or ``no'') has been transferred outwards,
across the Schwarzschild black-hole horizon.
\item
Pollux can transmit the message in his logbook (``YES'' or ``NO'')
to distant observers by classical means such as pulses of
electromagnetic radiation; see Sec.~\ref{sec:Discussion} for further discussion.

\end{enumerate}
\noindent A few technical remarks are in order:
\begin{enumerate}
\item[ad 4b.]
Castor and Pollux's procedure can be
extended by having several sequences (labeled $i=1,\, \ldots\,,\, I$)
with each $N_i^{P}=N$ electrons emitted by Pollux
and $N_i^{C}=N/0$ muons emitted by Castor,
so that Castor's whole message is
(yes/no,\,yes/no,\, $\ldots$ \,,\,  yes/no) with $I$ entries.
\item[ad 4d.]
As the probability for getting a measurable kick of the exterior electron
is small (see Apps.~\ref{app:Final-particle-trajectories}
and \ref{app:Probability-amplitude-in-position-space}),
$N$ needs to be taken sufficiently large.
\item[ad 4e.]
It is, in principle, possible
that Castor's $N$ muons do not produce a recoil electron
($N$ is large but finite)
or that Pollux's detector records a false ``1''
(the detector may trigger even in the absence of a recoil electron),
so that the  read message is not error-free.
Castor and Pollux may, therefore, decide to use an error-correcting code
for the $I$-entries message mentioned in the first technical remark.
\end{enumerate}
All these technical issues
are engineering questions and need to be addressed.
Note that the whole experimental setup of Castor and Pollux
(with a muon factory, a linear accelerator for muons, an electron source,
a linear accelerator for electrons, and a detector for electrons)
may have a substantial mass, but still very much less than
the black-hole mass $M$ which can be made arbitrarily large
(at least, in a \textit{Gedankenexperiment}).

%%\newpage%%tmp
\vspace*{0mm}
\section{Discussion}
\label{sec:Discussion}
\vspace*{0mm}

Heuristically, across-horizon Coulomb scattering
appears to be quite natural. Assume that a
nonrotating astrophysical black hole is initially neutral
and that, at a later moment, a particle of electric charge $Q$ falls in
radially, crossing the Schwarzschild horizon.
From a macroscopic point of view, this charged particle changes
the Schwarzschild black hole into a
Reissner--Nordstr{\o}m black hole of charge $Q$.
In other words, although the charged particle was ``swallowed" by the black hole,
knowledge of its charge $Q$ has
not disappeared for the region outside the black-hole horizon and
can still influence the outside-charges.
From a microscopic point of view, this Coulomb interaction happens due to virtual-photon exchange.

The across-horizon effect from electron-muon Coulomb scattering
happens due to a nonlocal correlation between the electron
and the muon. This correlation is described
by the Feynman propagator. The across-horizon effect
resembles, in fact,
%the state evolution with time of two
%causally disconnected localized detectors in Minkowski spacetime
the state evolution with time of two
causally-disconnected localized detectors in Minkowski spacetime %%FRK
interacting with each other through a local
relativistic quantum field, where the emerging entanglement
has been calculated for a simple interaction
model by Reznik \textit{et al.}~\cite{Reznik-etal-2005}.
The extraction of nonclassical correlations from the
quantum vacuum to particle detectors has been called
``entanglement harvesting'' in the modern literature
(cf.  Ref.~\cite{Pozas-Kerstjens-Martin-Martinez2016}
and references therein).

Let us elaborate on the analogy between our
across-horizon scattering (Fig.~\ref{fig:2})
and the setup of Reznik \textit{et al.},
which has two spatially separated detectors
in Minkowski spacetime
interacting via  a relativistic scalar field
(see Fig.~1 of Ref.~\cite{Reznik-etal-2005}).
The muon inside the black-hole horizon and the electron
outside
%can be considered as a pair of causally-disconnected ``detectors.''
can be considered as a pair of causally-disconnected ``detectors.''  %%FRK
%The local interaction between our ``detectors,''
The interaction between our ``detectors,''   %%FRK
i.e., the muon and the electron, is mediated by the photon field
and essentially lasts for a finite time interval
of order $\alpha\hbar/E_\text{CM}$. As a consequence of the
virtual-photon exchange between parts of this relativistic quantum system, the initial state $|\mu,p;e,k\rangle$ evolves unitarily with time
and has a finite overlap with
a particular final state $|\mu,p';e,k'\rangle$.
The analogy is, of course, not perfect.
For example, our ``detectors'' possess
infinitely many energy levels, whereas
the setup of Ref.~\cite{Reznik-etal-2005} considered
two-energy-level detectors (more realistic systems
have been considered in, e.g., Ref.~\cite{Pozas-Kerstjens-Martin-Martinez2016}).

Returning to our discussion of Coulomb scattering
across the Schwarzschild black-hole horizon,
the momentum exchange $q$ between inside-particles and outside-particles
is a measurable observable (Sec.~\ref{sec:Scattering}
and App.~\ref{app:Probability-amplitude-in-position-space}).
An outside-observer can measure, in principle,
the corresponding momentum change of the outside-particle,
but not the charge of the inside-particle or its initial momentum.
It appears that the across-horizon-Coulomb-scattering
effect can be employed to encode a message, which can
be sent by an inside-observer to an outside-observer
(Sec.~\ref{sec:Gedankenexperiment}).

It is sometimes said that
``an event horizon is the boundary in spacetime between events that can
communicate with distant observers and events that cannot''
(quote from Ref.~\cite{Schutz2009}, Sec.~11.3).
In view of the results of the present article, this statement
needs to be refined (the
spacetime manifold remains a classical concept):
``an event horizon is the boundary in spacetime
between events that can communicate with distant observers
by classical means and events that cannot.''
Nature is, of course, not classical ($\hbar \ne 0$)
and this fact allows for a potential breach of the event horizon
as defined by the second statement. We have indeed
shown that quantum scattering allows, in principle, for the transfer of
information from inside the black-hole event horizon to outside.

In this article, we have focussed on across-horizon Coulomb scattering
for astrophysical black holes, but similar effects may occur
in analogue systems~\cite{Weinfurtner-etal2011,Euve-etal2016}.

%%\newpage%%tmp
\vspace*{0mm}
\begin{appendix}
\section{Initial-particle trajectories}
\label{app:Initial-particle-trajectories}
\vspace*{0mm}

In this appendix, we give an example of initial-particle
trajectories which have
their closest approach near the Schwarzschild black-hole horizon, with
one particle (muon) inside the horizon
and the other particle (electron) outside the horizon.
Both particles are considered to be ultrarelativistic
and $c$ is set to unity. But, before we discuss these initial-particle
trajectories, we have a remark on the meaning
of the black-hole event horizon.

Generally speaking,
the event horizon is a global notion in the sense that it
depends on the observer's entire geodesic history and the large-scale
structure of spacetime~\cite{HawkingEllis1973}.
The black-hole event horizon, in particular, is associated
with a null hypersurface which corresponds to the boundary between null
rays that cannot come out and those that can.
This horizon surface can be projected locally on the
local inertial coordinate system (LICS) of the near-horizon region.
By the Einstein Equivalence Principle, the physics in the
freely-falling LICS
is the same as that of flat Minkowski spacetime
and nothing distinguishes locally the projected horizon
surface, only that part of this surface
coincides with part of the light-cone.

With appropriate local inertial coordinates
$\{T,\,X,\,Y,\,Z\}$ near the Schwarzschild horizon (Fig.~\ref{fig:2})
and a very large black-hole mass ${M}$, the horizon coincides
with a disk-like patch of
the $(Y,\,Z)$-plane centered around $(Y,\,Z)=(0,\,0)$ and
at position
\beqa
\label{eq:Xhor}
X_\text{hor}&=& T \,.
\eeqa
For the initial wave packages, we can take the following trajectories:
\bsubeqs\label{eq:XYmu-XYe-initial}
\beqa
\label{eq:XYe-initial}
\big(X,\,Y,\,Z\big)_{e,\,\text{in}} &\sim&
\big(-\sqrt{1/3}\;T + d_{0}- d_{\mu},\,\sqrt{2/3}\;T,\,0\big)\,,
\\[2mm]
\label{eq:XYmu-initial}
\big(X,\,Y,\,Z\big)_{\mu,\,\text{in}} &\sim&
\big(T - d_{\mu},\,0,\,0\big) \,,
\eeqa
\esubeqs
where the nonzero particle masses have been neglected for the
velocities.
The distance between the two wave packages is given by
\beqa\label{eq:d}
d(T)  &=&  \sqrt{(X_{\mu}-X_{e})^2+(Y_{\mu}-Y_{e})^2+(Z_{\mu}-Z_{e})^2} \,,
\eeqa
and the trajectories \eqref{eq:XYmu-XYe-initial}   %%FRK
result in having $d(0)=d_0$.

\begin{figure}[t]
\vspace*{5mm}
\includegraphics[scale=0.25]{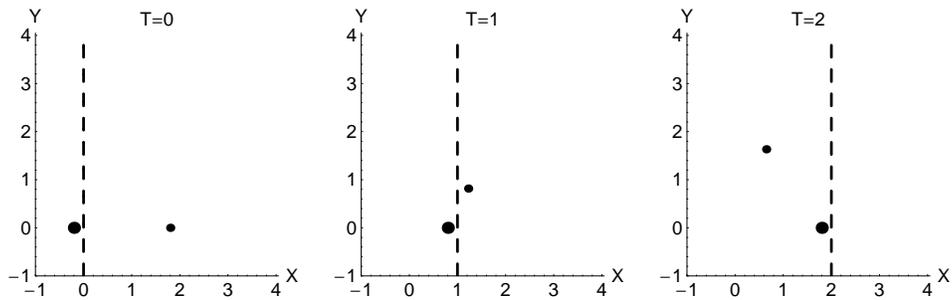} %%FRK
%Fig4-v7.eps=Fig4-v5.eps
%{BH-Gedankenexperiment-Fig4-v471.eps}
\vspace*{0mm}
\caption{Initial wave-package trajectories in the $(X,Y)$-plane
from \eqref{eq:XYe-initial} and \eqref{eq:XYmu-initial},
without dispersion and scattering.
The muon wave package (large dot) has increasing values
of $X$ and a constant value of $Y$.
The electron wave package (small dot)
has decreasing values of $X$ and increasing values of $Y$.
The projected black-hole horizon from \eqref{eq:Xhor}
is shown as the dashed line.
With an arbitrary length unit, the parameters in \eqref{eq:XYmu-XYe-initial}
are chosen as $\{d_{0},\,d_{\mu}\}=\{2,\, 1/5\}$.
The minimum separation $d_\text{min} \approx 0.92$
occurs at $T_\text{min}=1$.
For a length scale of the order of \eqref{eq:d-optimal},
significant Coulomb scattering occurs, as discussed in
Sec.~\ref{sec:Gedankenexperiment}
and App.~\ref{app:Probability-amplitude-in-position-space}.
}
\label{fig:4}
\end{figure}

An example of these trajectories
is given by Fig.~\ref{fig:4},
%% [using ],   %%FRK
where, at the moment
of closest approach ($T=1$), the electron is still outside the
horizon, $X_{e,\,\text{in}}(1) > X_\text{hor}(1)$,
while the muon is always inside,
$X_{\mu,\,\text{in}}(T) < X_\text{hor}(T)$.
Dispersion and scattering effects are neglected.
The particular initial-particle trajectories
of Fig.~\ref{fig:4} for a length scale of the order
of \eqref{eq:d-optimal}
give the momentum $k$ of the electron and
the momentum $p$ of the muon,
where $k$ and $p$ enter the amplitude \eqref{eq:2-2-scattering}.

Remark that, if the muon initially has a velocity with
nonzero $Y$ and $Z$ components, the horizon at position
\eqref{eq:Xhor} runs away from that particle
and the situation sketched in the middle panel of Fig.~\ref{fig:4}
does not occur. The coordinate $X$ lies, to leading order,
in the radial direction of the black-hole
spacetime (Fig.~\ref{fig:2}). The conclusion is, thus,
that the initial ultrarelativistic muon
(momentum $p$ in Fig.~\ref{fig:3}) must have, to
high precision, an outward radial motion.

%%\newpage%%tmp
\section{Final-particle trajectories}
\label{app:Final-particle-trajectories}
\vspace*{0mm}

In this appendix, we give a simplified
discussion of how Coulomb scattering may
affect the motion of the initial particles.
In App.~\ref{app:Probability-amplitude-in-position-space},
we give a detailed calculation
of the quantum scattering process in position space.

For the initial trajectories
of App.~\ref{app:Initial-particle-trajectories},
the large-angle electron-muon Coulomb scattering
probability is significant if the minimal distance in Fig.~\ref{fig:4}
is of order $\alpha\,\hbar c/E_\text{CM}$,
based on the estimate \eqref{eq:d-optimal} in terms of
the center-of-mass scattering energy $E_\text{CM}$. In that case,
there is a nonvanishing probability that the electron recoils.
An example of such a recoil trajectory is as follows:
\beqa
\label{eq:XYe-final-bad}
\big(X,\,Y,\,Z\big)_{e,\,\text{out}}\,\Big|^{(T>1)}
&\sim&
\big( X_{e,\,\text{out},\,1}+ \sqrt{1/2}\;(T-1),\,Y_{e,\,\text{out},\,1}
+ \sqrt{1/2}\;(T-1),\,0\big) \,,
\eeqa
with a matching muon trajectory from energy-momentum conservation
and a length unit equal to \eqref{eq:d-optimal}.
The constants $(X_{e,\,\text{out},\,1},\,Y_{e,\,\text{out},\,1})$
in \eqref{eq:XYe-final-bad}
correspond approximately to the position
of the initial electron at $T=1$ in Fig.~\ref{fig:4}.
A recoil electron with trajectory \eqref{eq:XYe-final-bad}
is, however, rapidly overrun by the horizon at position \eqref{eq:Xhor}.
Remark that, for a genuine scattering process,
we should only consider the
electron in \eqref{eq:XYe-final-bad} at $T\gg 1$,
but we have simplified the discussion somewhat by taking
$T> 1$.

\begin{figure}[t]
\vspace*{0mm}
\includegraphics[scale=0.25]{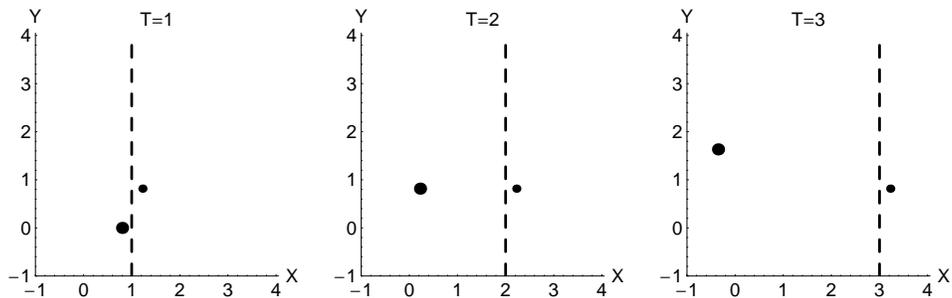} %%FRK
%Fig5-v7.eps=Fig5-v5.eps
%{BH-Gedankenexperiment-Fig5-v471.eps}
\vspace*{0mm}
\caption{Final wave-package trajectories in the $(X,Y)$-plane
from \eqref{eq:XYe-final} and \eqref{eq:XYmu-final},
with Coulomb scattering taking place at $T\sim 1$.
The constants $(X_{e,\,\text{out},\,1},\,Y_{e,\,\text{out},\,1})$
and $(X_{\mu,\,\text{out},\,1},\,Y_{\mu,\,\text{out},\,1})$
have been taken equal to the $T=1$ positions of the
initial-particle trajectories \eqref{eq:XYe-initial}
and \eqref{eq:XYmu-initial},
%as shown in Fig.~\ref{fig:4} for $T \leq 1$
%and a length scale of the order of \eqref{eq:d-optimal}.
as shown in Fig.~\ref{fig:4} for $T \leq 1$
and with a length scale of the order of \eqref{eq:d-optimal}.  %%FRK
}
\label{fig:5}
\end{figure}

Consider, next, a different trajectory of the recoil electron,
\bsubeqs\label{eq:XYmu-XYe-final}
\beqa
\label{eq:XYe-final}
\hspace*{-5mm}
\big(X,\,Y,\,Z\big)_{e,\,\text{out}}\,\Big|^{(T>1)}
&\sim&
\big( X_{e,\,\text{out},\,1}+ (T-1),\,Y_{e,\,\text{out},\,1},\,0\big) \,,
\\[2mm]
\label{eq:XYmu-final}
\hspace*{-5mm}
\big(X,\,Y,\,Z\big)_{\mu,\,\text{out}}\,\Big|^{(T>1)}
&\sim&
\big( X_{\mu,\,\text{out},\,1}  -\sqrt{1/3}\;(T-1),\,Y_{\mu,\,\text{out},\,1} +\sqrt{2/3}\;(T-1),\,0\big)\,,
\eeqa
\esubeqs
with constants $(X_{e,\,\text{out},\,1},\,Y_{e,\,\text{out},\,1})$
and $(X_{\mu,\,\text{out},\,1},\,Y_{\mu,\,\text{out},\,1})$
corresponding approximately  to the positions
of the initial particles at $T=1$ in Fig.~\ref{fig:4}.
Now, the final electron has a velocity purely in the $X$ direction
and the final electron stays outside the horizon,
provided \mbox{$X_{e,\,\text{out},\,1}>1$.}
Figure~\ref{fig:5} shows these final trajectories,
again using \eqref{eq:d-optimal} as the length unit.
The particular final-particle trajectories
of Fig.~\ref{fig:5} for a length scale of the order
of \eqref{eq:d-optimal}
give the momentum $k'$ of the electron and
the momentum $p'$ of the muon,
where $k'$ and $p'$ enter the amplitude \eqref{eq:2-2-scattering}.

The final recoil electron of \eqref{eq:XYe-final}
is special in that its $Y$ and
$Z$ velocity components are exactly zero.
Let us estimate which changes in that velocity direction
are allowed if we demand
that the final ultrarelativistic
recoil electron stays outside the black-hole horizon, at least,
over the Minkowski patch considered.

The curvature scale is given by
$r_S = 2\,G_{N}\,M/c^2 \approx 3\;\text{km}\;(M/M_\text{Sun})$
and the Minkowski patch has approximately that size.
If we now assume that, just after the scattering moment $T\sim 1$,
the recoil electron is outside the horizon by a distance
$\widetilde{d} \sim X_{e,\,\text{out},\,1}-1
\sim \alpha\,\hbar c/E_\text{CM}
%\approx 1.5\times 10^{-18}\;\text{m}\;(\text{GeV}/E_\text{CM})$
 \approx 1.4\times 10^{-18}\;\text{m}\;(\text{GeV}/E_\text{CM})$  %%FRK
in the $X$ direction,
then we find that the electron  trajectory can differ from the
trajectory \eqref{eq:XYe-final} by a small angle $\delta$
which is of order
\beqa
\label{eq:delta-estimate}
\delta
&\sim&
\sqrt{2\,\widetilde{d}/r_S} \sim
\sqrt{\alpha}\,E_P/\sqrt{E_\text{CM}\;M\, c^2} \,,
\eeqa
with Planck energy
$E_P \equiv  \sqrt{\hbar c^5/G_{N}}
\approx 1.22 \times 10^{19}\;\text{GeV}
\approx 2.18 \times 10^{-5}\;\text{g}$.
Hence, the allowed solid angle $\delta^2$ of the recoil electron
(momentum $k'$ in Fig.~\ref{fig:3})
is extremely small: $\delta^2 \sim 10^{-36}$ for
$E_\text{CM}\sim 10^{15}\,\text{GeV}$
and $M \sim M_\text{Sun} \sim 10^{38}\,E_P$.

A related issue is that the recoil electron
must have a sufficiently large energy
[a large enough gamma factor $E(k')/(m_e\, c^2)$],
so that the electron is not overrun by the horizon \eqref{eq:Xhor}
before the electron reaches the edge of the Minkowski patch at $X\sim r_S$.
Taking that the recoil electron at $T\sim 1$
is outside the horizon by a distance
$\widetilde{d} \sim \alpha\,\hbar c/E_\text{CM}$
in the $X$ direction,
and assuming a velocity purely in the $X$ direction
and an energy of order $E_\text{CM}/2$, we get the following
estimate for the required center-of-mass scattering energy:
\beqa
\label{eq:E-CM-bound}
E_\text{CM}
&\gtrsim&
4 \, \frac{1}{\alpha}\,\frac{m_e\,c^2}{E_P}\,\frac{M\,c^2}{E_P}\;m_e\,c^2\,,
\eeqa
where $m_e \approx 0.511\,\text{MeV}/c^2$ is the electron mass
and $M$ the black-hole mass.
With $M \sim M_\text{Sun}$, the required scattering
energy is $E_\text{CM} \gtrsim 10^{15}\,\text{GeV}$,
which is large but still less than the Planck energy $E_P$.

Considering the curved spacetime manifold of the
Schwarzschild black hole, it can be verified that
\eqref{eq:E-CM-bound} corresponds to the condition
for radial escape of the recoil electron if its initial energy
is of order $E_\text{CM}$ in the relevant local inertial frame
near the horizon and its initial position is at
a proper distance \eqref{eq:d-optimal} outside the horizon.
As the recoil electron escapes towards infinity
(point $i^{+}$ of Fig.~\ref{fig:3}), its energy
is red-shifted; cf. Sec.~25.4 in
Ref.~\cite{MisnerThorneWheeler1973}.

The conclusion is that both estimates \eqref{eq:delta-estimate}
and \eqref{eq:E-CM-bound} prefer a relatively small value of the
black-hole mass $M$.
Practical considerations, on the other hand, may favor
a relatively large value of $M$, as discussed in
the point 4a and
the last paragraph of
Sec.~\ref{sec:Gedankenexperiment}.

%%\newpage%%tmp
\section{Scattering probability amplitude in position space}
\label{app:Probability-amplitude-in-position-space}

In this appendix,
we discuss the elastic scattering process \eqref{eq:2-2-scattering}
in position space. Here, we will directly follow Feynman's
seminal paper~\cite{Feynman1949}.
An alternative calculation with auxiliary momentum variables
has been presented in App.~C of an earlier version of the
present article~\cite{EmelyanovKlinkhamer2017}.
Throughout this appendix, we use
the Cartesian coordinates $(X)^\mu =(T,\,X,\,Y,\,Z)^\mu$ with
Minkowski metric $\eta_{\mu\nu}$.
These coordinates appeared as local inertial
coordinates in Fig.~\ref{fig:2} and were already used in
Apps.~\ref{app:Initial-particle-trajectories}
and \ref{app:Final-particle-trajectories}.

\begin{figure}[t]
\vspace*{0mm}
\includegraphics[scale=0.9]{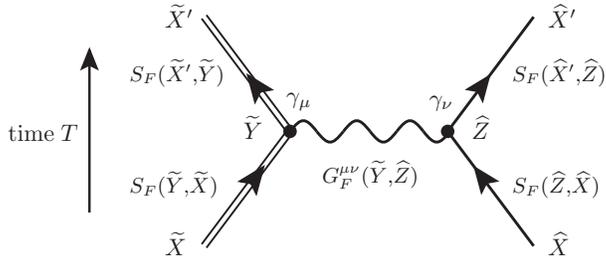}
%{ps-cs-v7003.eps}
%{BH-Gedankenexperiment-Fig6-v8.eps}
%{ps-cs-v60691.eps}
\vspace*{0mm}
\caption{Position-space Feynman diagram for elastic
electron-muon scattering
(adapted from Fig.~1 of Ref.~\cite{Feynman1949}).
The external muon is denoted by a double line and
the external electron by a single line, where
the flow of negative electric charge is indicated by an arrow.
The Feynman propagator of the photon is denoted by a wavy line.
The spacetime positions of the two interaction vertices (dots)
are to be integrated over. The corresponding probability
amplitude is given by \eqref{eq:Feynman-amplitude}.
}
\label{fig:6}
\end{figure}

We start from the original Feynman amplitude for 2-2 scattering
in position space, which is given by Eq.~(4) in Ref.~\cite{Feynman1949}.
Making some minor changes in notation (cf. Fig.~\ref{fig:6}),
we have the following tree-level probability amplitude
for an electron to propagate from $\widehat{X}$
to $\widehat{X}'$ and a muon from $\widetilde{X}$ to $\widetilde{X}'$\,:
\beqa\label{eq:Feynman-amplitude}
\hspace*{-7mm}
A(\widetilde{X}';\widehat{X}'|\widetilde{X};\widehat{X})\,\Big|^\text{tree} &=&
-e^2 \int{d^4\widetilde{Y}\, d^4\widehat{Z}}
\nonumber\\[2mm]
\hspace*{-7mm}
&& \times
S_{F}(\widetilde{X}',\widetilde{Y})\,\gamma_\mu\, S_{F}(\widetilde{Y},\widetilde{X})
\; G_{F}^{\,\mu\nu}(\widetilde{Y},\widehat{Z})\;
S_{F}(\widehat{X}',\widehat{Z})\,\gamma_\nu \,S_{F}(\widehat{Z},\widehat{X})\,,
\eeqa
where $S_{F}$ corresponds to the Feynman propagator of the
fermionic field (spinor indices are suppressed everywhere)
and $G_{F}^{\,\mu\nu}$ is the Feynman
propagator of the photon field in the Feynman gauge,
\beqa\label{eq:G-F}
G_{F}^{\,\mu\nu}(X',X) &=& -\frac{1}{4\pi^2}\,
\frac{\eta^{\mu\nu}}{\big[X'{-}X\big]^2 {-} i\epsilon}\,,
\eeqa
with the shorthand notation $\big[X'{-}X\big]^2 \equiv
\big(X'^\rho {-} X^\rho\big)\,\big(X'_\rho {-} X_\rho\big)$.
The Feynman propagator $G_{F}^{\,\mu\nu}(X',X)$
is also nonvanishing for spacelike separated points $X'$ an​d $X$.
In
%%the   %%FRK
contrast,
the retarded Green's function of the photon field
(relevant to classical processes)
only has a nonvanishing support for light-like separated
points $X'$ and $X$,
\beqa\label{eq:G-R}
G_{R}^{\,\mu\nu}(X',X) &=& \frac{i}{2\pi}\,\eta^{\mu\nu}\,
\theta\Big(X^{\prime 0} {-} X^0\Big)\,
\delta\Big(\big[X'{-}X\big]^2\Big)\,.
\eeqa

\begin{figure}[t]
\vspace*{0mm}
\includegraphics[scale=0.85]{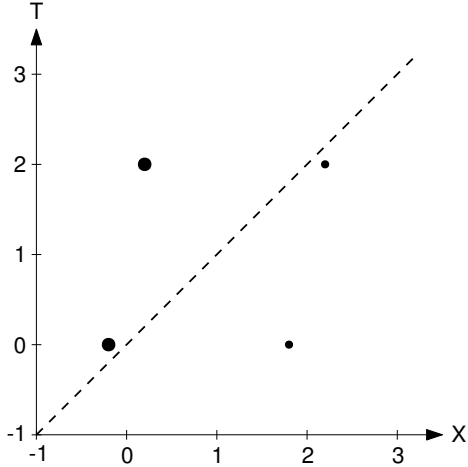} %%FRK
%{em-s-v7003.eps}
%{em-s-v6073.eps}
\vspace*{0mm}
\caption{Spacetime diagram with the positions
\eqref{eq:initial-and-final-positions}  %%FRK
of the initial and final particles for the electron-muon scattering
process considered (Fig.~\ref{fig:6}),
where the large dot corresponds to the muon $\mu^{-}$
and the small dot to the electron $e^{-}$.
Only the $T$ and $X$ coordinates are shown.
The initial positions at $T=0$ correspond to the colliding wave packages
of Fig.~\ref{fig:4}
and the final positions at $T=2$ correspond to the separating wave packages
of Fig.~\ref{fig:5}.
The dashed curve corresponds to a slice of the projected
black-hole horizon,
$\{X,\,Y,\,Z\}_\text{horizon\:at\:time\:$T$}= \{T,\,Y,\,Z\}$.
}
\label{fig:7}
\end{figure}

Now take the following $\{T,\,X,\,Y,\,Z\}$
values for the spacetime positions
of the initial and final particles:
\bsubeqs\label{eq:initial-and-final-positions}
\beqa
\widehat{X}_\text{electron}&=&
\Big\{0,\,+9/5,\,0,\,0\Big\}
\,,\\[2mm]
\widetilde{X}_\text{muon}&=&
\Big\{0,\,-1/5,\,0,\,0\Big\}
\,,\\[2mm]
\widehat{X}^{\prime}_\text{electron}&=&
\Big\{2,\,14/5-\sqrt{1/3},\,\sqrt{2/3},\,0\Big\}
\,,\\[2mm]
\widetilde{X}^{\prime}_\text{muon}&=&
\Big\{2,\,4/5-\sqrt{1/3},\,\sqrt{2/3},\,0\Big\}\,,
\eeqa
\esubeqs
where, for definiteness, we have used the values from
Figs.~\ref{fig:4} and \ref{fig:5}
for a length scale of the order of \eqref{eq:d-optimal}.
The positions \eqref{eq:initial-and-final-positions}  %%FRK
are shown in Fig.~\ref{fig:7}, together
with the projected black-hole horizon (dashed curve) as discussed in
Fig.~\ref{fig:2} and
App.~\ref{app:Initial-particle-trajectories}.
The leading-order probability amplitude for the electron-muon
scattering considered in Figs.~\ref{fig:6} and \ref{fig:7} is then
given by the scalar product of the initial and final wave
packages of the particles with the tree-level
amplitude \eqref{eq:Feynman-amplitude}.

Specializing to the initial and final positions
of Fig.~\ref{fig:7},
the tree-level amplitude \eqref{eq:Feynman-amplitude}
is nonzero,  as the Feynman propagator $G_{F}$
also has support outside the lightcone
and is nonzero even for spacelike separations.
The corresponding probability is then nonvanishing and gives
the recoil electron needed for the \textit{Gedankenexperiment}
of Sec.~\ref{sec:Gedankenexperiment}.

%%\newpage
Let us end this appendix by discussing the elastic
electron-muon scattering process (Fig.~\ref{fig:6})
in Minkowski spacetime \textit{per se},
leaving aside the application to black holes.
In order to simplify the discussion,  we temporarily
restrict to $1+1$ spacetime dimensions (coordinates $T$ and $X$).
We take it that the colliding particles
have positions at $T=0$ as given by Fig.~\ref{fig:7}
and that a recoil electron at $T=2$ is detected at
the position shown by Fig.~\ref{fig:7}. Specifically, there
is then a right-moving recoil electron detected
\emph{outside} the right-moving lightcone of the initial muon at $T=0$.
Observe that the process does not involve
faster-than-$c$ propagation of light~\cite{DrummondHathrell1980},
which, incidentally, may still be consistent with
causality~\cite{Liberati-etal-2002,KlinkhamerSchreck-2011}.
Rather, there is the propagation of a \emph{virtual}
photon over a \emph{spacelike} distance,
as discussed in Sec.~\ref{sec:Introduction}.

Note that, due to the nonzero mass of the electron, the
right-moving recoil electron of Fig.~\ref{fig:7} will ultimately
get \emph{inside} the right-moving lightcone of the initial muon.
As mentioned already in App.~\ref{app:Final-particle-trajectories},
we should, for a genuine scattering process,
only consider final particles at $T\gg 2$
and initial particles at $T\ll 0$.
For this genuine scattering process, the time-ordering
of the initial and final particles is invariant
under the action of proper orthochronous Lorentz transformations.

\begin{figure}[t]
\vspace*{0mm}
\includegraphics[scale=0.90]{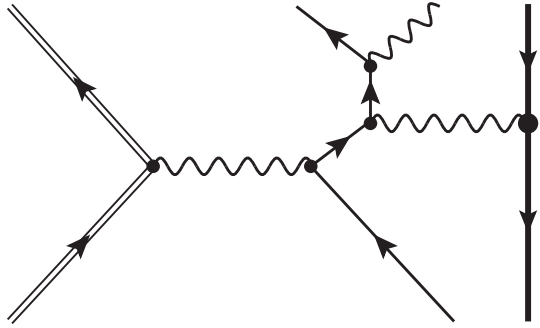}  %%FRK
%{em-photon-nucleus-v70037.eps}  %%heavy line --> final fig8
%{em-photon-nucleus-v70095.eps} %%crossed circle left
%{em-photon-nucleus-v70030.eps}  %%crossed circle right
\vspace*{-0mm}
\caption{Position-space Feynman diagram for electron-muon scattering
(Fig.~\ref{fig:6}) with Brems\-strahlung from the
scattered electron as it is deflected by a nucleus
(symbolically shown by the heavy line on the right).
For the Bremsstrahlung of the scattered electron,
there are, in general, two more Feynman diagrams to consider;
cf. Fig.~6.21 of Ref.~\cite{ItzyksonZuber1980}.  %%FRK
The spacetime positions of the five interaction vertices (dots)
are to be integrated over.
}
\label{fig:8}
\end{figure}

Now consider the following modification of
this elastic scattering process:
a subprocess is added which transfers the information from
the ``detection'' of a right-moving electron at $T=2$ in Fig.~\ref{fig:7}
to a right-moving electromagnetic signal.
This right-moving electromagnetic signal does stay outside
the right-moving lightcone of the initial muon.
At the microscopic level, the subprocess can be realized as
Bremsstrahlung~\cite{ItzyksonZuber1980}
from the scattered electron as it is deflected by a nucleus;
see Fig.~\ref{fig:8}.
There is, then, a small but nonvanishing probability to get
a right-moving hard photon which stays outside
the right-moving lightcone of the initial muon.
(Causality holds as long as the right-moving hard photon is
inside the right-moving lightcone of the initial electron or nucleus;
see also the discussion below Eq.~(3.8) and Fig.~3
in Ref.~\cite{Dickinson-etal2014}.)
It will be a challenge to design a Minkowski-spacetime experiment
to demonstrate this QED effect with a final hard photon
outside the lightcone of the initial muon.

\end{appendix}

%%FRK

\end{document}